# A History-Guided Regional Partitioning Evolutionary Optimization for Solving the Flexible Job Shop Problem with Limited Multi-load Automated Guided Vehicles

Feige Liu, Chao Lu, Xin Li

*Abstract*—In a flexible job shop environment, using Automated Guided Vehicles (AGVs) to transport jobs and process materials is an important way to promote the intelligence of the workshop. Compared with single-load AGVs, multi-load AGVs can improve AGV utilization, reduce path conflicts, etc. Therefore, this study proposes a history-guided regional partitioning algorithm (HRPEO) for the flexible job shop scheduling problem with limited multi-load AGVs (FJSPMA). First, the encoding and decoding rules are designed according to the characteristics of multi-load AGVs, and then the initialization rule based on the branch and bound method is used to generate the initial population. Second, to prevent the algorithm from falling into a local optimum, the algorithm adopts a regional partitioning strategy. This strategy divides the solution space into multiple regions and measures the potential of the regions. After that, cluster the regions into multiple clusters in each iteration, and selects individuals for evolutionary search based on the set of clusters. Third, a local search strategy is designed to improve the exploitation ability of the algorithm, which uses a greedy approach to optimize machines selection and transportation sequence according to the characteristics of FJSPMA. Finally, a large number of experiments are carried out on the benchmarks to test the performance of the algorithm. Compared with multiple advanced algorithms, the results show that the HRPEO has a better advantage in solving FJSPMA.

## I. INTRODUCTION

With the development of production automation, traditional manufacturing has gradually transformed and upgraded to intelligent manufacturing[1]. Among them, the automated logistics and transportation of the workshop has largely promoted the optimization and upgrading of the production system. Thanks to the development of technologies such as artificial intelligence, automatic guided vehicles (AGVs) were initially used for express logistics and the transportation of warehouse goods[2]. Due to the flexibility and efficiency of AGV, it has gradually become the basic equipment for workshop logistics in recent years[3]. Especially with the increasing competition in the manufacturing industry, in order to meet the diverse needs of customers, the production model has gradually shifted to a flexible job shop scheduling problem (FJSP) with multiple varieties and small batches[4]. This production model is widely used in solid-state capacitor production[5], automobile production (chassis, engine and other parts production and assembly)[6], etc. Using AGVs for transportation can not only perform transportation tasks flexibly, but also achieve uninterrupted production and improve production efficiency.

The FJSP with transportation resource constraints is an NP-hard problem[7]. Different from the problem of directly providing jobs transportation schedule, the problem of considering AGVs to transport jobs is more complicated because AGVs are resources that need to be allocated and arranged in scheduling. The transportation path planning problem of AGVs is embedded in FJSP, that is, the transportation tasks on AGVs and the processing tasks on processing machines are coupled. Therefore, when studying the FJSP with the transportation resource, it is more in line with the actual production environment to consider AGVs as a resource[8]. In manufacturing systems, AGVs as a resource that needs to be allocated and scheduled, are often limited. To improve the utilization rate of AGVs, it is possible to consider further upgrading the AGV from being able to load only one job to being able to load multiple jobs. We call this type of AGV a multi-load AGV. At present, the application field of multi-load AGVs is mainly in automatic storage and retrieval systems (AS/RS)[9]. Compared with single-load AGVs, multi-load AGV systems can simultaneously meet the goals of reducing the number of occupied AGVs and minimizing conflicts[10]. Based on the advantages of multi-load AGVs and their application in warehousing systems, we find that multi-load AGVs also have great potential in optimizing intelligent production systems. Therefore, this paper studies a flexible job shop scheduling problem considering limited multi-load AGVs (FJSPMA).

To solve FJSPMA, this study proposed a history-guided region partitioning algorithm (HRPEO). The goal is to minimize the maximum completion time. The main

This work was supported by National Natural Science Foundation of China under Grant Nos. 52175490 and 51805495. (*Corresponding author: Chao Lu*).

Feige Liu and Chao Lu are with School of Computer Science, China University of Geosciences, Wuhan, China. (E-mail: liufeige3@163.com; luchao@cug.edu.cn).
Xin Li is with Department of Mathematics and Information Technology, The Education University of Hong Kong, HongKong, China. (E-mail: stephenli@eduhk.hk)

contributions are as follows: (1) An optimization model of FJSPMA is established, and encoding and decoding rules are designed. (2) An initialization strategy based on branch and bound and heuristic rules is designed. (3) An evolutionary framework for search region partitioning is adopted to guide the subsequent search process based on the information of the historical solutions. (4) A local search strategy is designed based on the characteristics of multi-load AGV scheduling.

The rest of this paper is organized as follows: Section II is an overview of literature review. Section III introduces the problem. Section IV provides a detailed description of the HRPEO. Section V tests the influence of parameters and the effectiveness of various strategies in the HRPEO. Section VI summarizes the conclusions of this article and discusses future development directions.

## II. LITERATURE REVIEW

### A. FJSPMA

At present, the AGVs in FJSP can be roughly divided into unlimited AGVs and limited AGVs according to the number transportation resources. That is, the former considers that there are countless AGVs available for transportation tasks in the production environment. Sufficient transportation resources will not lead to the situation that the operation waits for AGV after it is completed. The makepsan of this type of problem is often not affected by the selection of AGVs and the sequence of transportation tasks on AGV. Jiang et al.[11] studied an energy-conscious flexible job shop scheduling problem considering transportation time and deterioration effect. They only considered the transportation time, without considering the allocation of transportation resources and the scheduling of transportation tasks. However, in the actual production process, the available AGVs are often limited. Andy[12] proposed two different constraint programming formulations for the first time for a FJSP with limited transport robot. Arash et al.[13] developed a Mixed Integer Linear Programming (MILP) and a parallel heuristic (PTSDBH) for FJSP considering conflict-free AGV routing. PTSDBH first forms multiple incomplete solutions and then refines each solution based on the best value. Li et al.[14] proposed an efficient optimization algorithm to solve the flexible job shop scheduling problem with crane transportation processes (CFJSP). The crane can be used as a single-load AGV with a fixed transport route. In addition to optimizing makespan, Yao et al.[15] considered the energy-saving and designed a DQN-based memetic algorithm to solve this problem. Dalila et al.[16] also studied an energy-efficient job shop scheduling problem with transport resources considering speed adjustable resources. They developed a bi-objective mixed-integer linear programming model for the problem. Further, Li et al.[17] developed a hybrid deep Q-network for dynamic multi-objective flexible job shop scheduling problem with insufficient transportation resources. The DQN maked agent learn to select the appropriate rule at each decision point. Gao et al.[18] propose a cloud-edge combined digital twin (DT) flexible job shop scheduling (CE-DTFJSS) framework to address the flexible job shop scheduling problem with limited AGVs under abnormal disturbances.

The above studies all consider the use of single-load AGVs for the transportation of jobs or processing materials in FJSP. However, the scheduling problem of multi-load AGVs is mainly about path planning, task scheduling, etc. Hu et al.[19] studied the multi-load AGV scheduling problem in network logistics systems and considered the situation of transportation path conflicts. They developed an appointment schedule that can be used to prevent AGV collisions and deadlocks. Lin et al.[10] proposed a task scheduling optimization method for multi-load AGVs-based systems, and they considered incremental task sizes, changeable maximum load of AGVs and the number of available AGVs. Dang et al.[20] studied the scheduling of heterogeneous multi-load AGVs with battery constraints. They proposed a large neighborhood search method to generate AGV scheduling solutions with low lateness cost and low AGV transportation cost. Yan et al.[21] used an advanced form of Petri Nets to develop a model of a multi-load AGV system. This paper found that the efficiency of the AGV system can be improved by increasing the load capacity of AGV, but the effectiveness of such an approach will decrease when the load capacity increases above a certain value. The above studies show that multi-load AGVs with appropriate load limits can improve the transportation efficiency of AGVs, and numerical verification has been carried out on logistics scheduling examples. However, in the field of production scheduling, there are few cases where multi-load AGV scheduling is embedded in FJSP, so it is necessary to study FJSPMA.

### B. HRPEO

For general evolutionary algorithms, since the algorithm always prefers to converge in a region that is easy to search, it is difficult to find the global optimum by regenerating the initial solutions and repeating the evolution when converging to the local optimum. Therefore, in order to change the convergence preference of the algorithm, some algorithms will record the current optimal solution, and after reinitialization, the evolution process tends to move away from the local optimal solution that has been found. This approach implicitly divides the whole into two different search regions. It is similar to the niching technique. The current representative niching methods include preselection[22], cropping[23] and sharing[24]. These methods all search by implicitly or explicitly dividing the population into multiple sub-populations. Wang et al.[25] developed a penalty-based differential evolution algorithm and constructed a dynamic penalty radius. During selection, the region within the penalty radius of the recorded elite solution will be penalized. Wang et al.[26] proposed an automatic niching technique based on the affinity propagation clustering. This strategy can automatically divide the clusters. Xia et al.[27] proposed an evolutionary algorithm based on reinforcement learning, and used Q-learning to search the position and clustering subspace. Liu et al.[28] proposed a two-layer clustering morphological differential evolution

(DLCSDE) algorithm. The first layer of clustering divides the population into multiple sub-populations for positioning, and the second layer searches for local and global optimality missed by the first layer during clustering. Wang et al.[29] tested several advanced niche-based evolutionary algorithms and found that although these algorithms avoid repeating similar evolutionary processes, they cannot avoid the individuals in the region that has already converged solution need to being ignored. Therefore, they proposed a history-guided hill search to avoid searching for previously obtained solutions.

This search strategy is widely used in solving multimodal problems. Although the goal of solving FJSPMA with evolutionary algorithms is to find the global optimal scheduling solution, it is not necessary to find all the optimal solutions, but this strategy based on niching technology helps prevent the algorithm from falling into the local optimum. Therefore, it is worth studying the use of niching technology to optimize the evolutionary algorithm for solving FJSPMA.

III. PROBLEM DESCRIPTION

*A. Problem statement*

Fig 1 shows a workshop layout with multi-load AGVs. In simple terms, FJSPMA is a combination of FJSP and Vehicle Routing Problem (VRP). In more detail, the FJSPMA can be described as $N$ jobs need to be processed on $K$ machines, and each job $J_i$ ($i$ is the index of the jobs) contains $S_i$ operations. Each operation $O_{ij}$ ($j$ is the index of the operations of job $J_i$) can be processed on at least one machine. In addition, there is a raw material warehouse (MW) for storing initial processing raw material, and a product warehouse (PW) for storing completed processing jobs. There are $RN$ multi-load AGVs used for transport jobs between processing machines and warehouses. $A$ is the upper limit of the capacity of multi-load AGVs, and the size of each job is 1. The goal of FJSPMA is to minimize the maximum completion time. FJSPMA satisfies the constraints of FJSP and also needs to satisfy the following constraints:

1) All jobs and AGVs are ready at the MW from time zero.
2) The maximum load of each AGV during transportation shall not exceed the upper capacity of the AGV.
3) It is assumed that the loading and unloading time of the AGV is taken into account in the processing time.
4) The buffer of each machine is infinite, and if many jobs arrive in the buffer of a machine, the order of process on the machine is not constrained by the order in which the jobs arrive.
5) When starting processing, jobs need to be transported from MW to the machine, and after completing processing, jobs need to be transported to PW.
6) The path conflicts caused by AGVs transporting jobs are not considered.

In brief, the FJSPMA can be subdivided into four sub-problems:
1) Determine the processing order of all operations for all jobs.
2) Select processing machines for each operation.
3) Select a AGV machine for each operation
4) Determine the sequence of loading and unloading tasks of all jobs on each AGV.

Each operation of the job has two transportation processes before starting processing, namely, the AGV goes to the current location of the job to load the job, the AGV transports the job to the machine where the next operation is located or to PW, as shown in Fig 2. Fig 2 illustrates a partial transport route of an AGV with the capacity of 2, where the green squares correspond to loading tasks, the purple squares correspond to unloading tasks, and the circular nodes correspond to processing tasks of operations. For a single-load AGV, loading and unloading tasks are performed alternately, while for a multi-load AGV, loading or unloading tasks can be performed continuously.

*B. Mathematical Model of FJSPMA*

Yao et al.[30] proposed a novel mixed integer linear programming (MILP) model for the flexible job-shop scheduling problem with limited single-load AGVs. This study considers multi-load AGVs, so for an operation, the loading and unloading tasks on the AGV are not necessarily continuous. This section mainly introduces the constraints on multi-load AGVs:

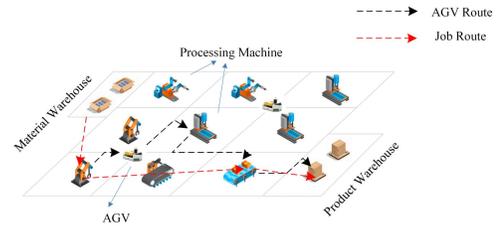

**Fig. 1.** Workshop layout with multi-load AGVs.

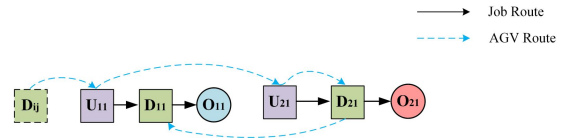

**Fig. 2.** A partial transport route of an AGV.

**Indices:**

$j, j'$: indices for operations, and include the last operation (the processing time is 0) of each job which is the operation in PW. $j, j' = 1, 2, 3, \cdots, N$, where $N$ is the total number of operations.

$t$: to distinguish the type of transport task, $t = 1$ represents a loading task, and $t = 0$ represents an unloading task.

$r$: index of AGVs, $r = 1, 2, \cdots, R$.

$k$; index of machine, $k = 0, 2, \cdots, K$, $k = 0$ means MW.

**Parameters:**

$N_f$: set of the first operation of all jobs.

$V$: a very large positive number.

$A$: The upper limit of the capacity of multi-load AGVs.

$ST_{jt}$: the start time of transport task for operation $j$, the type of task is $t$.

$TT_{k_1 k_2}$: the transport time from machine $k_1$ to machine $k_2$.

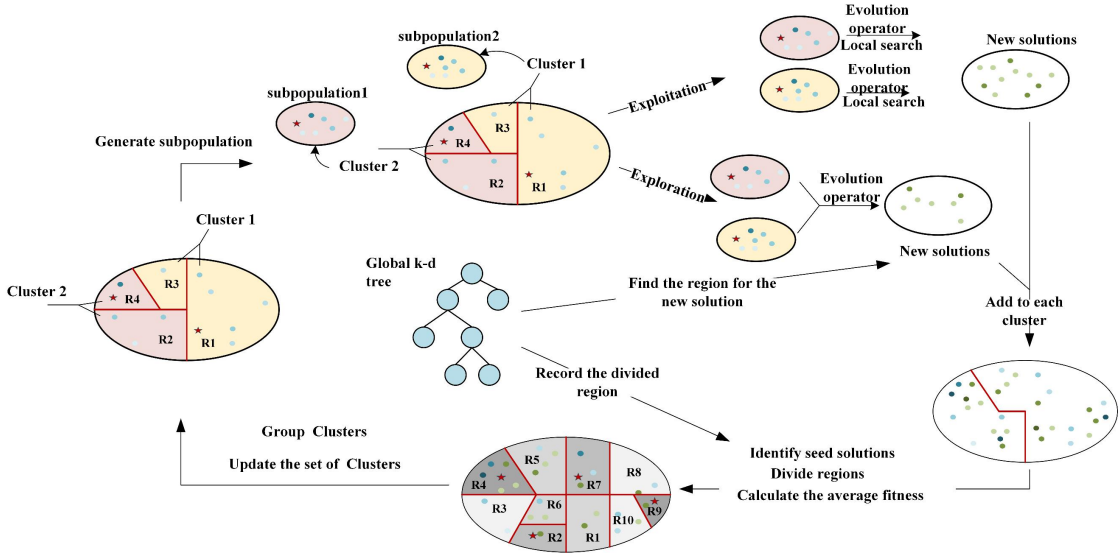

**Fig. 3.** The general process of the HRPEO.

$SO_j$: the start time of processing task for operation $j$.
$PT_j$: the processing time of operation $j$.
**Decision variables:**
$x_{jrt}$: if AGV $r$ is selected to process the loading task of operation $j$ or the unloading task of operation $j$, $x_{jrt} = 1$; otherwise, $x_{jrt} = 0$;
$y_{jk}$: if operation $j$ is processed on machine $k$. $\eta_j$: if operation $j$ have transport tasks.
$\gamma_{jj'k}$: if operation $j'$ be processed after operation $j$ on machine $k$;
$\mu_{jtj't'r}$: if the transport task $t$ of operation $j$ is performed precedes the transport task $t'$ of operation $j'$ on AGV $r$, $\mu_{jtj't'r} = 1$; otherwise, $\mu_{jtj't'r} = 0$;
Minimize $C_{max}$:
$$C_{max} \geq SO_j + Pt_j, \forall j \epsilon N \quad (1)$$
(2) Constrain that an operation $j$ with transport tasks must have two transport tasks: loading and unloading.
$$\sum_t x_{jrt} = 2 * \eta_j, \forall j \epsilon N, \forall r \epsilon R \quad (2)$$
For multi-load AGVs, continuous loading and unloading tasks are allowed without exceeding the capacity limit of AGV. The following constraints are about subproblems 3 and 4.
(3) indicates that for any transport task on a multi-load AGV, starting from any $x_{jrt}$ tasks, in the sequential $2 * A$ transport tasks, the number of loading and unloading tasks is $A$.
$$(\mu_{j_1 t_1 j_2 t_2 r} + \mu_{j_2 t_2 j_3 t_3 r} + \cdots + \mu_{j_{2A-1} t_{2A-1} j_{2A} t_{2A} r} - 2A) \cdot V + x_{j_1 r t_1} + x_{j_2 r t_2} + \cdots + x_{j_{2A} r t_{2A}} \leq 2A ,$$
$$\forall r \epsilon R, \forall t_{1 \text{ to } 2A} \epsilon \{0,1\},$$
$$\forall j_{1 \text{ to } 2A} \epsilon N, j_1 t_1 \neq j_2 t_2 \neq \cdots \neq j_{2A} t_{2A} \quad (3)$$
(4) to (7) constrain the AGV to only perform one transportation task at a time. For the loading task of operation $j$, the machine corresponding to this task is the machine where the previous processing stage of the job is located, denoted as $k_{pre}$. For the unloading task of operation $j$, the machine corresponding to this task is the processing machine arranged by operation $j$, denoted as $k_{pro}$.

$$ST_{j't'} \geq ST_{jt} + (1-t) \cdot TT_{k_{pre}k_{pro}} + t \cdot TT_{k''_{pro}k''_{pre}} - V \cdot (8 - \mu_{jtj't'r} - \mu_{j''t''jtr} - x_{jrt} - x_{j'rt'} - x_{j''rt''} - \varphi_{jk_{pro}} - \varphi_{j-1k_{pre}} - t'' \cdot \varphi_{j''k''_{pro}} - (1-t'') \cdot \varphi_{j''-1k''_{pre}}),$$
$$\forall r \epsilon R, \forall t, t', t'' \epsilon \{0,1\}, \forall j, j'' \epsilon N, \forall j' \epsilon N - N_f, \forall k_{pre}, k_{pro}, k''_{pro}, k''_{pre} \epsilon K, j''t'' \neq jt \neq j't' \quad (4)$$
$$ST_{j't'} \geq ST_{jt} + (1-t) \cdot TT_{k_{pre}k_{pro}} + t \cdot TT_{0k_{pre}} - V \cdot (8 - \mu_{jtj't'r} - \mu_{j''1jtr} - x_{jrt} - x_{j'rt'} - x_{j''r1} - \varphi_{jk_{pro}} - \varphi_{j-1k_{pre}}),$$
$$\forall r \epsilon R, \forall t, t' \epsilon \{0,1\}, \forall j, j'' \epsilon N, \forall j' \epsilon N_f,$$
$$\forall k_{pre}, k_{pro} \epsilon K, j''1 \neq jt \neq j't' \quad (5)$$
$$\mu_{jtj't'r} + \mu_{j't'jtr} = 1,$$
$$\forall r \epsilon R, \forall t, t' \epsilon \{0,1\}, \forall j, j' \epsilon N, jt \neq j't' \quad (6)$$
$$\mu_{jtj''1r} + \mu_{j''1jtr} = 1,$$
$$\forall r \epsilon R, \forall t, t'' \epsilon \{0,1\}, \forall j, j'' \epsilon N, j''t'' \neq jt \quad (7)$$
(8) to (10) relate the transfer task to the processing task. Especially for the unloading task, the start time of the unloading task is affected by the completion time of the previous stage operation of the job.
$$SO_j \geq ST_{j1} + TT_{k_{pre}k_{pro}} - V \cdot (3 - \eta_j - \varphi_{jk_{pro}} - \varphi_{j-1k_{pre}}), \forall j \epsilon N - N_f, \forall k_{pre}, k_{pro} \epsilon K \quad (8)$$
$$SO_j \geq SO_{j-1} + PT_{j-1} - V \cdot \eta_j, \forall j \epsilon N - N_f \quad (9)$$
$$ST_{j1} + V \cdot (1 - \eta_j) \geq SO_{j-1} + PT_{j-1}, \forall j \epsilon N - N_f \quad (10)$$
(11) to (13) are common constraints of the FJSP problem, regarding subproblems 1 and 2.
$$\sum_k y_{jk} = 1, \forall j \epsilon N \quad (11)$$
$$\gamma_{jj'k} + \gamma_{j'jk} \leq 1, \forall j, j' \epsilon N, \forall k \epsilon K, j \neq j' \quad (12)$$
$$SO_{j'} \geq SO_j + PT_j - V \cdot (3 - y_{jk} - y_{j'k} - \gamma_{jj'k}),$$
$$\forall j, j' \epsilon N, \forall k \epsilon K, j \neq j' \quad (13)$$

IV. ALGORITHM DESCRIPTION

*A. The framework of the HRPEO*

Before introducing the algorithm, we will introduce region, cluster and subpopulation which will be used later. Region is a multidimensional interval specified by the problem encoding. If a solution is in the region, then the value of each dimension is within the interval of the corresponding dimension. Cluster refers to the collection of regions, which is a larger multidimensional interval. Subpopulation corresponds to each cluster one by one, and the solutions of subpopulation are all within the range of the corresponding cluster.

To minimize the completion time, this paper proposes an HRPEO. In HRPEO, a global k-d Tree is used to record the regions divided during the iteration process. And the historical solutions of each region also will be recorded. The core of HRPEO is to select appropriate individuals for optimization based on the metrics of different regions, and balance the exploitation and exploration capabilities of the algorithm. The contents of HRPEO are as follows: 1) The decoding method combined with heuristic rules is used to calculate the fitness. And to generate an initial population with uniform distribution and better individuals, we borrowed the branch and bound method to generate the initial population. 2) To determine the future search trends, a regional division strategy is designed. In each iteration, update the set of clusters, identify the seed solutions, and then divide the regions. After that, cluster all the regions to form a new set of clusters. 3) Carry out exploitation and exploration. Exploitation is to generate subpopulations corresponding to clusters and perform evolution operations on individuals in a same subpopulation. Exploration is to select two different clusters based on roulette and evolve the individuals in their regions. 4) To mention the search efficiency, a local search strategy is designed. According to the metric of each region, that is, the average fitness value of all solutions in people and, select appropriate individuals for local search.

The general process of the HRPEO is shown in Fig 3. Each point represents a searched solution. The darker the color, the better the solution. The yellow and red areas represent different clusters. Each cluster generates a corresponding sub-population. After a series of search operations, the new solutions are represented by a green point. Then the seed solutions are determined in each cluster, and the region is divided. The search potential of the region can be evaluated by the average fitness of all solutions in the region. The grayer the region, the more likely it is to find a local optimal solution. Then the neighboring regions are aggregated into a cluster based on the potential of the region. For more details, see Algorithm 1.

### B. Encoding and Decoding

According to the four sub-problems of the problem, this paper adopts a three-layer encoding with a transport tasks list. The decoding method is combined with a certain loading and unloading sequence to obtain a scheduling solution for FJSPMA.

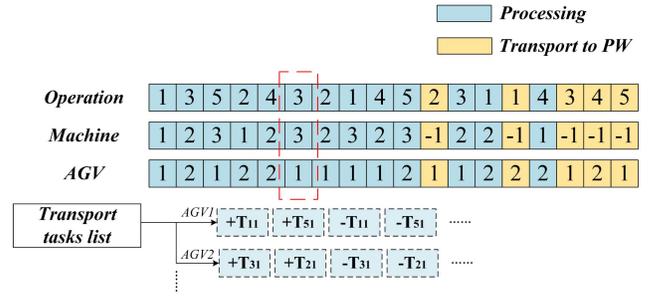

Fig. 4. encoding of the FJSPMA.

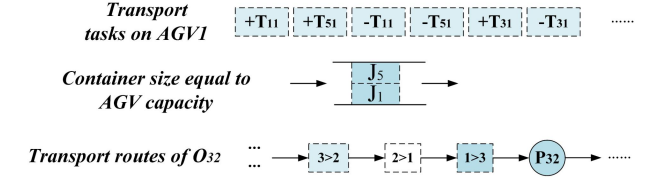

Fig. 5. Part of the transport path on the AGV.

| Algorithm 1. HRPEO |
| --- |
| 1. **Input**: size of initial population $N \cdot N1$, number of exploratory solutions $N2$, $\alpha$ to determine the range of the interval; |
| 2. **Output**: the best solution $X_{best}$ |
| 3. // Initialization |
| 4. $S \leftarrow$ Initialize the population $(N \cdot N1)$; |
| 5. a collection of clusters $CS \leftarrow \emptyset$, subpopulations $SP \leftarrow \emptyset$; |
| 6. Determine the value range of each bit of the solution vector, then obtain the range of the initial solution space $R$; |
| 7. a collection of regions $RS \leftarrow \emptyset$; $RS \leftarrow RS \cup R$ |
| 8. **While** $t < Iteration\ End\ Time$ **do** |
| 9. //Exploitation |
| 10. $SP \leftarrow$ Generate subpopulation $(CS, SP, RS)$; |
| 11. $SP, RS \leftarrow$ Each subpopulation performs crossover mutation and local search $(SP, RS)$; |
| 12. //Exploration |
| 13. **For** $i = 1\ to\ N2$ **do** |
| 14. **If** $|CS| > 1$ **then** |
| 15. Select two clusters $C1$ and $C2$; |
| 16. $X_{new}$ = Generate a solution $(C1, C2)$; |
| 17. $CS, RS \leftarrow$ Update the solution set corresponding to the region $(CS, RS, X_{new})$; |
| 18. **Else** |
| 19. $X_{new}$ = Generate a solution randomly; |
| 20. **End** |
| 21. **End** |
| 22. //Update $CS$ and $RS$ |
| 23. **For** $i = 1\ to\ |CS|$ **do** |
| 24. $U \leftarrow$ Identify seed solutions (the solution set of $CS_i$); |
| 25. Update $X_{best}$; |
| 26. $RS \leftarrow$ Divide regions $(RS, U)$; |
| 27. **End** |
| 28. $CS \leftarrow$ Group Clusters $(RS)$; |
| 29. **End** |

As shown in Fig 4, the first layer is the coding of all operation processing sequences, and the blue box represents the operation processing task. It is worth noting that, for the convenience of decoding, it is assumed that the operation in the yellow box is performed in the PW, the processing time is set to 0, and the processing machine is marked as -1. The second layer is the coding of the machine, which corresponds to the operation coding and indicates the processing machine where the operation is located. The third layer is the coding of the AGVs, which is similar to the machine coding and indicates the index of the AGV selected for the operation. The transport task list records the order of transport tasks on each AGV. When the task number is +, it is a loading task, and when it is −, it is an unloading task.

When decoding, a container with a capacity equal to the AGVs capacity is required. According to the correspondence between AGV and operation in the code, the transportation tasks on each AGV are listed in the processing order of the operations. Each operation has two transportation tasks, the loading task is $+T_{ij}$, and the unloading task is $-T_{ij}$. In particular, if two consecutive operations of a job are processed on the same machine, the transport tasks of the latter are not taken into account. If it is a loading task, a job block is stored in the container, and if it is an unloading task, the corresponding job block is removed from the container. Initially, if the container is not full, the loading task is processed, and if the container is full, all the jobs in the container are unloaded one by one. Fig 5 gives an example of an AGV with a capacity of 2, and gives the transportation route of job 3 in the red box in Fig 4 on the AGV. According to the above rules, the order of transportation tasks on each AGV can be obtained. After that, you can decode them one by one according to the processing order of the operations. The specific steps are as follows:

Step1: According to the operation processing order, get the operation $O_{ij}$. Then, determine the assigned processing machine and AGV, the current position of the job.

Step2: According to the order of transportation tasks on AGV, get the position of AGV. Process the load task of $O_{ij}$. The start time of $+T_{ij}$ is the completion time of the previous task in the AGV. The end time is the start time plus the transportation time from the current location of the AGV to the current location of the job.

Step3: Continue to traverse the AGV task list. If the next task is $-T_{ij}$, calculate the start time of the task $-T_{ij}$. At this time, compare the end time of the previous task of the AGV with the completion time of the operation $O_{ij-1}$ (if the operation is the first operation, the completion time is 0) on its current machine, and select the larger one as the start time of $-T_{ij}$. The completion time is the start time plus the transportation time from the current machine of the job to the machine assigned to $O_{ij}$. Otherwise turn to Step 1.

Step4: Determine when to start processing an operation. The larger of the completion time of $-T_{ij}$ in AGV and the completion time of an operation on the scheduled machine that before $O_{ij}$ is used as the start processing time of $O_{ij}$. Easy to get the completion time of $O_{ij}$.

Step5: If the last operation in the operation list is traversed, the maximum completion time is output, otherwise go to Step 1.

Based on the above decoding strategy, Fig 6 shows the Gantt chart corresponding to the encoding in Fig 4, where boxes of the same color represent processing tasks or transportation task of the same job. Boxes with two colors on the AGV represent two jobs on the AGV during the current transportation process.

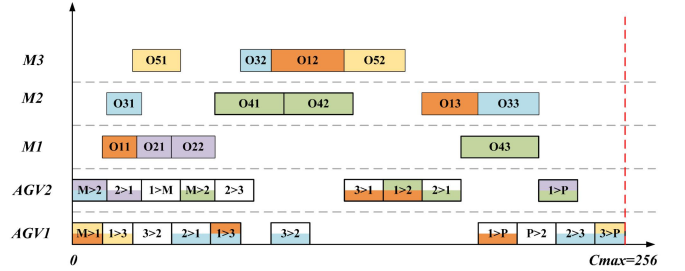

**Fig. 6.** The Gantt chart for a solution.

### C. The seed solutions

The purpose of dividing the solution space into multiple regions is to continuously search for unsearched regions during the search process. At the same time, local searches can be performed in the searched regions for in-depth search. Compared with the classic evolutionary algorithm, this search strategy divides similar solutions into a region and explores unknown regions at the same time, which can effectively improve the exploration ability of the algorithm.

The principle of dividing regions in this paper is to divide the local optimal solutions with large similarity differences into different regions. Such local optimal solutions are called seed solutions. Therefore, the update of clusters in the iterative process is to first identify the seed solutions within the cluster range, divide the regions according to the seed solutions, and then perform clustering to generate multiple clusters. When clustering, make sure that any two seed solutions are not in the same cluster. In order to implement the above method, this paper adopts the metric NBD mentioned in Preuss's clustering method[31]. This metric is based on the similarity of solutions in a set and the ranking of the fitness of the solutions in the set. The calculation formula of this metric is as follows:

$$NBD(X_i, I) = \begin{cases} \infty & , if\ B_{X_i,I} = \emptyset \\ \min_{X_k \in B_{X_i,I}} |X_i - X_k| & , otherwise, \end{cases}$$
(10)

$B_{X_i,I}$ is the set of solutions that are better than $X_i$ in the solution set $I$, and $|X_i - X_k|$ denotes the Euclidean distance between solutions $X_i$ and $X_k$.

According to the calculation formula of NBD, it can be inferred that if the solutions in the set are evenly distributed, most solutions can find better solutions in the closer neighborhood, so the smaller the value of NBD. If the value of NBD is large, it means that the fitness of the corresponding solution is better and the difference with other better solutions is greater. In order to verify that the relationship between the solutions to this problem conforms to the above inference, we conducted two experiments. Experiment 1 randomly generates 1000 solutions and calculates the NBD value. The histogram of NBD of all solutions is shown in Fig 7 (a). Experiment 2 is the histogram of NBD of all solutions in a cluster during the iteration process of HRPEO, see Fig 7 (b). In Fig 7, each outlier in the distribution is a local optimal solution, so these outliers can be regarded as seed solutions. In order to detect outliers in the distribution, this paper defines solutions with an NBD value greater than $\mu + \alpha \cdot \sigma$ as the seed solutions of the distribution. Where $\mu$ is the mean of the distribution and $\sigma$ is the standard deviation. In addition, it is necessary to filter out similar solutions, so that the seed solutions corresponding to the cluster can be obtained.

| Algorithm 2. Divide regions ($RS, U$) |
| --- |
| **Input:** a collection of regions $RS$, seed solution set $U$; |
| **Output:** updated $RS$; |
| 1. // Initialization |
| 2. **For** $i = 1\ to\ |U|$ **do** |
| 3.    sign←true; |
| 4.    **For** $j = 1\ to\ |U|$ **do** |
| 5.       **If** $X_i \ne X_j$ **then** |
| 6.          Identify the dimensions with the greatest difference $k$; |
| 7.          $R_i$ ←Find the region where $X_i$ is located in $RS$; |
| 8.          $R_j$ ←Find the region where $X_j$ is located in $RS$; |
| 9.          **If** $R_i$ and $R_j$ overlap in the range of the $k$th dimension **then** |
| 10.            //Half the region |
| 11.            $R_{i1}, R_{i2}$ ←halve ($R_i$); |
| 12.            $(RS - R_i) \cup R_{i1} \cup R_{i2}$; |
| 13.          **End If** |
| 14.       **End If** |
| 15.    **End** |
| 16. **End** |

| Algorithm 3. Group Clusters ($RS$) |
| --- |
| **Input**: a collection of regions $RS$ |
| **Output**: a collection of clusters $CS$ |
| 1. $CS \leftarrow \emptyset$; $AED \leftarrow \emptyset$; |
| 2. Filter the area that must not have a corresponding solution ($RS$); |
| 3. $WTA$ ←Sort all regions in ascending order according to their mean fitness ($RS$); |
| 4. $idx \leftarrow 0$; |
| 5. **While** $|WTA| > 0$ **do** |
| 6.    $Cluster \leftarrow \emptyset$; |
| 7.    $Cluster \leftarrow$ Find the Neighborhood regions ($WTA, AED, WTA_i$); |
| 8.    $CS \leftarrow CS \cup Cluster$; |
| 9. **End** |

### D. Divide the Region and Construct clusters

The solution to the problem studied in this paper can be represented by a high-dimensional vector. Therefore, the region can be defined as a subspace parallel to the axis. In short, the region is to give a value range to each dimension of the high-dimensional vector. The cluster is a collection of regions. The k-d Tree is a suitable data structure that can be used for the storage and search of high-dimensional data. A hyperplane can be used to divide a region into two half-regions. For example, according to the k-th dimension division, the k-th dimension values of the points on the left side of the tree are all less than the specified value, and the nodes on the right are all greater than the specified value. Based on the structure of k-d Tree, we can continuously decompose the space into multiple subspaces. First, find the dimension with the largest variance among all seed solutions, and then divide the space into two regions according to this dimension. At the same time, distribute the original solutions to the two regions. The above method is shown in Fig 8.

Based on this region half-division method, the solutions in the cluster will be divided into different regions, so the seed solutions will also be assigned to different regions. When any two seed solutions in the cluster are not in adjacent regions, the region division can be stopped. Since the region is divided in half each time, the reference vector and region range of the nodes at the same height and position in the k-d tree are the same. Not only that, considering that the regions must be divided in each iteration and for each cluster, we construct a global k-d tree to record the regions divided during the entire iteration process. The solutions searched in each generation can also find the corresponding region in the k-d tree and copy it to the solution set of the region. The mean fitness of solution set in the region can be used as a standard to measure the potential of the region. See Algorithm 2 for details. In the subsequent iteration process, when it is necessary to locate the region for the seed solution in the global k-d Tree, if the traversed node has left and right subtrees, there is no need to repeatedly build the leaf node of the node. At this time, you only need to continue searching downward. When the termination condition is met, the region corresponding to the node where the index is located is the divided subregion. And during the traversal process, if the index traverses to the left subtree, the region corresponding to the right subtree needs to be added to the $RS$. Similarly, if the index is in the right subtree, the region on the left subtree needs to be added to the $RS$.

| Algorithm 4. Find the Neighborhood regions |
| --- |
| **Input:** $WTA, AED, WTA_i$ |
| **Output**: a collection of clusters $Cluster$ |
| 1. $Cluster \leftarrow Cluster \cup WTA_i$; |
| 2. $AED \leftarrow AED \cup WTA_i$ |
| 3. **For** $R \in$ Adjacent regions of $WTA_i$ **do** |
| 4.    **If** $R$ is not better than $WTA_i$ and $R \notin AED$ **then** |
| 5.       $Cluster \leftarrow$ Find the Neighborhood regions ($WTA, AED, R$); |
| 6.    **End If** |
| 7. **End** |

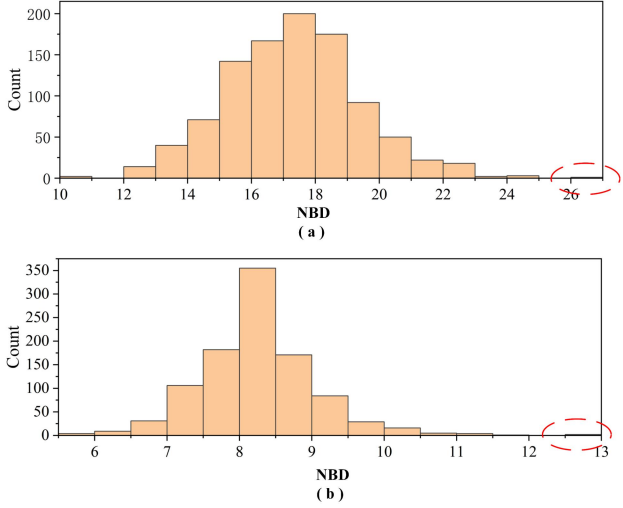

**Fig. 7.** A histogram of NBD of a random solution set (a) and A histogram of NBD of a solution set in a cluster (b).

After all the regions in the clusters are redivided, all the regions are stored in $RS$, and the cluster set is updated using the clustering method. As mentioned above, whenever the number of solutions in a region increases, the value of the region, that is, the mean of the solution set fitness, will be updated. Therefore, clusters can be constructed based on the mean fitness of each region. The clustering method is shown in Algorithm 3.

It should be explained that the purpose of filtering the regions in RS is to exclude the situation where the solutions in the region are definitely infeasible solutions. For example, there may be a dimension where High-Low=0, and there is no solution in this region. In addition, due to the constraints of the job sequence processing in the problem, the solutions in the region may be definitely infeasible solutions. Finding adjacent regions in Algorithm 3 is a recursive process, and AED is used to record the regions that have been assigned to a cluster. The process of finding adjacent regions to build clusters is shown in Algorithm 4.

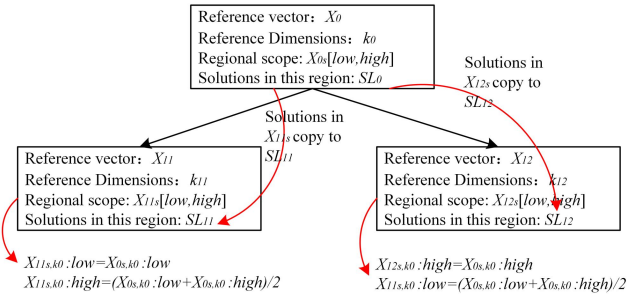

**Fig. 8.** The process of partitioning with hyperplanes in k-d Tree.

### E. Initialize the Population

Inspired by the parallel heuristic mentioned in Arash, this paper designs an initialization rule[13]. The process of generating solutions based on a decision tree to obtain a relatively promising initial solution while ensuring that the initial population distribution is relatively uniform. The decision tree is shown in Fig 9. The root node is the starting node, from which the decision tree is generated. The nodes in the second layer are the first operation of each job. The nodes in the third layer are the operation that can be selected after its parent node. Therefore, the processing order of the operations can be obtained based on the decision tree. For the selection of machines and AGVs, First Come First Served heuristic rule is used to determine. However, the solution space of the problem is large, so it is impossible to list all solutions and calculate the fitness of them. Therefore, pruning is performed when the decision tree generates the operation sequence. That is, the size of the initial population is set to $N \cdot N1$, and the subtree corresponding to each second-level root node ultimately retains only $N1$ branches. When constructing a certain layer of each subtree, if the leaf node is greater than $N1$, the fitness value of some solutions is calculated, and the branches with larger fitness values are subtracted.

In addition, in the subsequent iteration process, the size of each subpopulation is also $N \cdot N1$. Therefore, the size of the population is related to the instance. The larger the instance size, the larger the solution space and the larger the population size of the algorithm. This strategy can, to a certain extent, reduce the impact of instance size on algorithm performance.

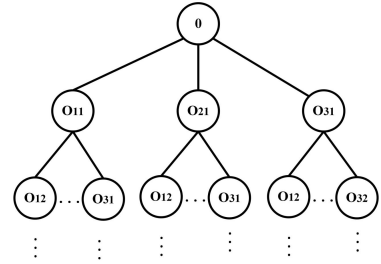

**Fig. 9.** The decision tree for generating initial population.

### F. Evolutionary Operators

Crossover and mutation operators are used in both Exploitation and Exploration. In Exploitation, the parents come from the same cluster, while in Exploration, the children are selected from different clusters.

As the iterative proceeds, the number of solutions in each region may increase, and accordingly, the number of solutions in each cluster may also increase. In the exploitation, we cannot use evolutionary operations on every solution, so we should give priority to solutions with greater potential. This paper considers selecting $N \cdot N1$ better individuals for each cluster as selected individuals to construct the subpopulation of this cluster. If the number of solutions in a cluster is greater than $N \cdot N1$, the first $N \cdot N1$ best solutions in the cluster are copied to the subpopulation. If the number of solutions in a cluster is less than $N \cdot N1$, all the solutions in the cluster are copied to the subpopulation. Then randomly generated solutions within the range corresponding to the cluster and add into the subpopulation until the number of individuals in the subpopulation is met. In Exploration, two clusters are selected by roulette according to the mean of fitness for each cluster, and two solutions are randomly selected from the clusters for evolution. The above process is repeated to generate a total of $N2$ solutions.

The above explains how to select parents in exploitation and exploration respectively. After selection, the operation encoding is performed to POX crossover and mutation operations[32, 33]. The machine and AGV encodings are subjected to PMX crossover and mutation operations[34]. In addition, the machine encoding may produce unworkable solutions after the change, so it needs to be corrected. The correction strategy is to randomly select a machine that is suitable for the operation.

In addition, all solutions generated by evolutionary operations must be stored in the global k-d tree.

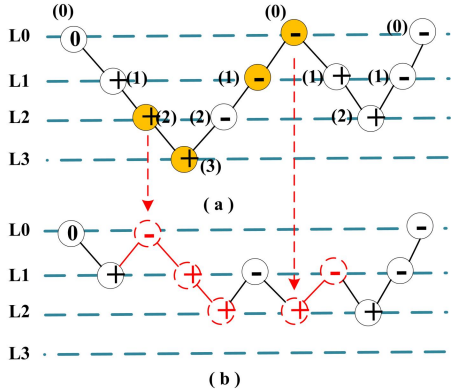

**Fig. 10.** The order of partial loading and unloading tasks on an AGV.

### G. Local Search

In exploitation, the crossover and mutation operator are used to let the offspring inherit the dominant gene fragments of the parent while randomly changing other fragments. This evolutionary operation may obtain a better solution, but it is random. Therefore, to improve the search efficiency of the exploitation, we use a search operator inspired by the feature of the problem. In this part, we use a greedy method, which has a strong local search ability, but takes more time[35]. To overcome this shortcoming, the local search strategy will only be executed when the fitness value of the solution is less than the mean fitness of the region to which it belongs.

The greedy search strategy is used to perform local search on machine codes and AGV codes. That is, the machine codes and AGV codes are traversed sequentially, and all optional machines and AGVs are tried. If the fitness of the solution is smaller after changing the machine and AGV, the original solution is replaced.

However, the search for the list of transport tasks on the AGV is more complicated. This is because the order of transport tasks on the AGV is affected by the upper limit of the AGV capacity. For example, when the upper limit of the AGV capacity is 2, the 3 consecutive transport tasks on the AGV cannot all be loading tasks. Therefore, we studied the order of loading and unloading tasks on multi-load AGVs. As shown in Fig 10, we exemplify the order of partial loading and unloading tasks on an AGV when the AGV capacity limit is 3. Assume that the loading task is represented by $+$, the unloading task is represented by $-$, and the code shows the jobs that each AGV needs to transport. Therefore, as long as the order of loading and unloading tasks on each AGV is known, the transportation path of the AGV can be known. In Fig 10, 0 represents the start, L0 represents the number of jobs on the current AGV is 0, L1 represents the number of jobs on the AGV is 1, and so on. As can be seen from Fig 10 (a), when it is a loading node, the number of layers where the node is located will increase relative to the previous node, otherwise, if it is an unloading node, it will decrease.

Due to the capacity constraint of the AGV, the position of all nodes will not exceed the given number of layers. Therefore, not all loading nodes can be changed to unloading nodes, and vice versa. Combining the above, we can find that for AGV with capacity 3, only the loading nodes at L2 and L3 can be transformed into unloading nodes, and the unloading nodes at L0 and L1 can be transformed into loading nodes. The transformation method is shown in Figure 10 (b). Node transformation is achieved by moving the node up or down one level. When a node is transformed, the subsequent nodes will also change accordingly. If the difference in the number of layers between a changed node and the node behind it is greater than 1, the node behind it must also move accordingly. Similarly, we can infer that for AGV with capacity 2, only the loading node at L2 and the unloading stage at L0 can be transformed. The rules of AGVs of other capacities can also be inferred in this way. It is worth noting that it is necessary to ensure that the last unloading node must be in L0.

Combined with the above rules, this paper uses local search for transportation tasks on AGV. It tries to transform each transformable node, and if the transformed solution is better, replace the original solution.

## V. EXPERIMENTAL RESULTS AND ANALYSIS

### A. Instance setting

This paper uses two benchmarks: FJSPT and EX[36], but the datasets do not consider the upper limit of the AGV capacity. Therefore, this paper sets the upper limit of the AGV capacity to 2 or 3. All compared algorithms are coded in Java. All experiments are run on a Microsoft Windows 11 operating system with 20 GB RAM of memory and a 3.20 GHz Intel(R) Core (TM) i5-10505 CPU. The iteration termination criterion of all the algorithms in all experiments is CPU time, and the running time of each algorithm is $N \times K \times A \times 10$ ms ($N$ is the number of jobs, $K$ is the number of machines, $A$ is the number of the multi-load AGVs). In all experiments, each algorithm is run independently for 20 times, and this paper uses $ARPD$ as an indicator to evaluate the performance of the algorithm.

$$ARPD = \frac{1}{R}\sum_{i=1}^{R}\frac{C_i - C_{best}}{C_{best}} \quad (11)$$

where $C_i$ is the makespan of solution found by current approach in $i$-th trial. $C_{best}$ is makespan of the current best solution in all trial times for instance. $R$ is the number of trial times.

TABLE I
THE ARPD OF HRPEO AND THE VARIANTS OF IT OF ALL INSTANCES

| Instance | HRPEO | HRPEO1 | HRPEO2 | HRPEO3 | HRPEO4 |
|---|---|---|---|---|---|
| 1 | **1.22414** | 1.96552 | 1.94828 | 7.24138 | 3.82759 |
| 2 | 2.43103 | 2.74138 | **2.32759** | 5.20690 | 3.44828 |
| 3 | **1.21429** | 1.30357 | 1.69643 | 9.33929 | 3.73214 |
| 4 | **0.72059** | 0.75000 | 1.05882 | 7.64706 | 3.72059 |
| 5 | **0.87302** | 0.88889 | 1.01587 | 5.71429 | 2.20635 |
| 6 | 1.28125 | 1.40625 | **1.25000** | 4.71875 | 2.00000 |
| 7 | 1.59677 | **1.58065** | 1.66129 | 6.85484 | 3.35484 |
| 8 | **1.41892** | 1.52703 | 1.51351 | 5.89189 | 2.45946 |
| 9 | **2.25862** | 2.55172 | 2.63793 | 10.51724 | 3.79310 |
| 10 | **2.12069** | 2.58621 | 2.65517 | 10.81034 | 3.74138 |
| 11 | **1.90909** | 2.30909 | 2.27273 | 9.89091 | 3.81818 |
| 12 | **1.46377** | 1.57971 | 1.75362 | 8.11594 | 4.46377 |
| 13 | **1.79167** | 2.45833 | 2.25000 | 8.22917 | 3.62500 |
| 14 | 1.52000 | 1.46000 | 1.68000 | 7.30000 | 2.54000 |
| 15 | **1.93478** | 2.30435 | 2.58696 | 12.78261 | 4.13043 |
| 16 | **2.00000** | 2.21667 | 2.01667 | 7.33333 | 3.35000 |
| 17 | **1.60563** | 1.91549 | 2.18310 | 11.87324 | 4.47887 |
| 18 | **1.60870** | 2.21739 | 2.65217 | 11.84058 | 4.92754 |
| 19 | **3.03030** | 3.28788 | 3.16667 | 11.43939 | 5.68182 |
| 20 | **2.45679** | 2.95062 | 3.16049 | 12.76543 | 5.65432 |
| 21 | **1.25882** | 1.61176 | 1.72941 | 6.55294 | 3.10588 |
| 22 | **1.84524** | 2.02381 | 2.21429 | 6.73810 | 2.82143 |
| 23 | **2.00000** | 2.66667 | 2.56790 | 6.72840 | 4.13580 |
| 24 | **1.80000** | 2.07368 | 2.56842 | 6.94737 | 4.62105 |
| 25 | **2.28169** | 2.61972 | 2.77465 | 10.90141 | 3.07042 |
| 26 | **2.05556** | 2.33333 | 2.38889 | 6.34722 | 3.76389 |
| 27 | **3.01493** | 3.37313 | 3.44776 | 3.76119 | 4.73134 |
| 28 | **2.73077** | 2.85897 | 2.88462 | 10.78205 | 4.46154 |
| 29 | **1.38346** | 1.66165 | 1.72180 | 4.61654 | 2.93985 |
| 30 | 2.29808 | 2.42308 | 2.31731 | 5.01923 | 3.18269 |
| 31 | 1.62281 | 1.83333 | 1.57018 | 5.05263 | 2.65789 |
| 32 | **1.42056** | 1.63551 | 1.46729 | 6.57009 | 2.98131 |
| 33 | 1.62651 | 1.80723 | 1.40964 | 5.13253 | 2.84337 |
| 34 | **0.98519** | 1.28148 | 1.30370 | 4.82963 | 2.60741 |
| 35 | **1.60377** | 1.80189 | 1.91509 | 5.77358 | 3.59434 |
| 36 | **0.36310** | 0.50000 | 0.72024 | 8.80952 | 1.58929 |
| 37 | **1.26357** | 1.49612 | 1.26357 | 5.13953 | 1.93798 |
| 38 | **0.88824** | 1.16471 | 1.19412 | 5.61176 | 2.68824 |

TABLE II
FRIEDMAN ANALYSIS OF EACH VARIANTS OF HRPEO

| Algorithm | mean | p-value |
|---|---|---|
| HRPEO/ HRPEO1 | 3.15582 | 0.016 |
| HRPEO/ HRPEO2 | 3.69993 | 0.00216 |
| HRPEO/ HRPEO3 | 10.48313 | 1.03266E-24 |
| HRPEO/ HRPEO4 | 7.87142 | 3.50646E-14 |

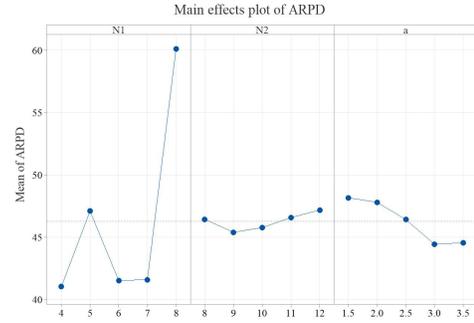

Fig. 11. The main effect plot for $ARPD$

### B. Experimental Parameters

The parameter settings have an impact on the performance of the algorithm. There are three parameters involved in the HRPEO, and Taguchi's experimental method is used to find the optimal combination of parameters in this section[37]. The levels of each parameter are as follows: $N1 = \{4,5,6,7,8\}$, $N2 = \{8,9,10,11,12\}$, $\alpha = \{1.5, 2.0, 2.5, 3.0, 3.5\}$. The main effect plot for $ARPD$ is shown in the Fig. 11. $N1$ has the greatest impact on the algorithm performance, followed by $\alpha$, and $N2$ has the least impact on the algorithm. $N1$ affects the search of HRPEO during the development process. If $N1$ is too large, there may be too many solutions that need to be optimized in the subpopulation, which will take too much time in the exploitation. If $N1$ becomes smaller, there will be a certain probability of forgetting potential solutions. $\alpha$ controls the number of the identified seed solutions. If it is too small, too many seed solutions will be identified, making the regional division too scattered. If $\alpha$ is too large, some seed solutions will be missed, making it difficult to divide the region. $N2$ controls the scope of global search to a certain extent. If $N2$ is too small, it is difficult to find solutions in unknown regions. If it is too large, it will interfere with the exploitation ability of the algorithm. Based on comprehensive evaluation, the optimal parameter configuration is as follows: $N1 = 6, N2 = 9, \alpha = 3.5$.

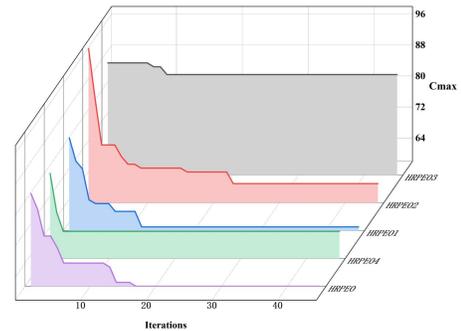

Fig. 12. The convergence curves of variants

TABLE III
THE ARPD OF HRPEO AND THE COMPARION ALGORITHM OF ALL INSTANCES

| Instance | DHNDE | DQNMMA | EDA_ACO_LS | EMOEA | LMEO | PSOSA | HRPEO |
|---|---|---|---|---|---|---|---|
| 1 | 9.862 | 10.793 | 4.207 | 4.207 | 9.121 | 10.741 | **1.431** |
| 2 | 9.638 | 10.914 | 3.793 | 3.793 | 8.948 | 11.190 | **2.690** |
| 3 | 11.607 | 10.946 | 3.571 | 3.571 | 10.321 | 11.018 | **1.839** |
| 4 | 9.618 | 11.412 | 3.426 | 3.426 | 9.647 | 9.809 | **0.765** |
| 5 | 10.825 | 10.444 | 3.984 | 3.984 | 9.651 | 10.937 | **1.048** |
| 6 | 9.250 | 10.469 | 4.344 | 4.344 | 8.766 | 10.266 | **1.453** |
| 7 | 10.839 | 10.774 | 4.339 | 4.339 | 9.452 | 11.048 | **1.887** |
| 8 | 10.486 | 11.054 | 4.811 | 4.811 | 10.892 | 11.257 | **1.473** |
| 9 | 13.207 | 16.052 | 5.190 | 5.190 | 11.948 | 14.224 | **2.776** |
| 10 | 14.224 | 15.000 | 5.534 | 5.534 | 12.121 | 14.483 | **2.603** |
| 11 | 14.636 | 16.545 | 6.291 | 6.291 | 13.945 | 15.891 | **2.018** |
| 12 | 15.377 | 16.435 | 5.986 | 5.986 | 15.522 | 15.884 | **1.623** |
| 13 | 9.792 | 11.396 | 4.667 | 4.667 | 8.333 | 10.438 | **2.167** |
| 14 | 8.240 | 11.260 | 3.860 | 3.860 | 7.540 | 9.820 | **1.780** |
| 15 | 10.717 | 12.500 | 5.435 | 5.435 | 9.957 | 11.587 | **2.065** |
| 16 | 10.483 | 10.933 | 4.833 | 4.833 | 9.700 | 10.950 | **2.300** |
| 17 | 10.803 | 11.042 | 6.141 | 6.141 | 10.465 | 12.634 | **2.282** |
| 18 | 11.884 | 11.928 | 6.754 | 6.754 | 10.449 | 12.942 | **2.391** |
| 19 | 13.485 | 13.667 | 8.636 | 8.636 | 12.530 | 14.848 | **3.348** |
| 20 | 13.568 | 14.951 | 7.198 | 7.198 | 14.086 | 15.025 | **3.333** |
| 21 | 10.847 | 11.353 | 6.000 | 6.000 | 10.165 | 12.118 | **1.741** |
| 22 | 11.560 | 11.298 | 7.833 | 7.833 | 11.179 | 11.988 | **2.321** |
| 23 | 13.173 | 13.778 | 8.728 | 8.728 | 11.926 | 13.605 | **2.840** |
| 24 | 12.726 | 13.000 | 7.674 | 7.674 | 13.179 | 13.189 | **2.537** |
| 25 | 10.690 | 11.380 | 5.169 | 5.169 | 10.000 | 11.352 | **2.606** |
| 26 | 10.222 | 10.833 | 4.556 | 4.556 | 9.778 | 11.264 | **2.417** |
| 27 | 11.104 | 13.254 | 7.313 | 7.313 | 11.209 | 12.388 | **3.403** |
| 28 | 11.872 | 13.167 | 6.513 | 6.513 | 12.679 | 14.000 | **2.756** |
| 29 | 8.226 | 7.338 | 5.421 | 5.421 | 3.805 | 9.263 | **1.774** |
| 30 | 8.144 | 6.865 | 6.596 | 6.596 | 3.904 | 8.788 | **2.558** |
| 31 | 7.053 | 7.447 | 6.404 | 6.404 | 3.237 | 9.202 | **1.781** |
| 32 | 8.673 | 6.308 | 5.991 | 5.991 | 4.710 | 8.972 | **1.701** |
| 33 | 7.277 | 6.229 | 5.386 | 5.386 | 3.928 | 8.337 | **1.639** |
| 34 | 6.222 | 6.111 | 4.563 | 4.563 | 2.867 | 6.956 | **1.259** |
| 35 | 7.868 | 7.283 | 7.151 | 7.151 | 4.264 | 9.481 | **1.943** |
| 36 | 4.750 | 4.982 | 3.613 | 3.613 | 2.452 | 6.274 | **0.375** |
| 37 | 5.395 | 5.504 | 1.922 | 1.922 | 2.597 | 6.876 | **1.225** |
| 38 | 5.835 | 6.035 | 5.059 | 5.059 | 3.353 | 7.759 | **1.135** |

*C. Effectiveness of Each Improvement Strategy of HRPEO*

To verify the effectiveness of each strategy in HRPEO, this section compares HRPEO with its variants. The details of each variant are as follows:

HRPEO1: the exploration in HRPEO is removed;
HRPEO2: the initialization strategy is removed and random initialization is adopted;
HRPEO3: the local search strategy is removed;
HRPEO4: the region division strategy is removed, and the GA framework is adopted while retaining other strategies, and the tournament selection strategy is adopted. The experimental results are shown in Table I. In addition, we used the Friedman analysis of variance method to analyze the data of each group at the 0.05 level, and the analysis results are shown in Table II. Fig. 12 illustrates the convergence curves of these variants on an example. From the experimental results, it can be seen that in most instances, HRPEO has a smaller ARPD value. Fig. 12 can well reflect the role of each strategy. HRPEO1 has a slightly weaker exploration ability than HRPEO and has not converged to the optimal solution. HRPEO2 uses random initialization, so the initial fitness value is relatively large. HRPEO3 lacks a search strategy based on problem characteristics and intelligent random search, so it is difficult to find the optimal solution. HRPEO4 lacks regional division and is easy to condense local optimality. HRPEO has achieved good results on 32 instances. However, in the instance 7 and 14, it is easier to find a better solution without exploration strategy. This is because the solution in the exploration process may take more time to perform regional division, thereby reducing the search efficiency. In the instance 2, 6, 31 and 33, the initialization method mentioned in this article will discard individuals with better potential during pruning, which causes the algorithm to fall into a local optimal state.

In summary, the local search strategy and the region division strategy have made the main contribution to the improvement of the algorithm performance. The initialization strategy and the exploration strategy have also improved the performance of the algorithm to a certain extent, but due to the influence of the instance, they do not work every time.

TABLE IV
FRIEDMAN ANALYSIS OF EACH COMPARION ALGORITHM

| Algorithm | mean | p-value |
|---|---|---|
| DHNDE/ HRPEO | 8.28351 | 2.51153E-15 |
| DQNMMA/ HRPEO | 10.19509 | 4.3796E-23 |
| EDA_ACO_LS/ RPEO | 3.50456 | 0.0096 |
| EMOEA/ HRPEO | 3.50456 | 0.0096 |
| LMEO/ HRPEO | 5.52234 | 7.0248E-7 |
| PSOSA/ HRPEO | 11.36328 | 1.33789E-28 |

*C. Effectiveness Comparisons to other algorithms*

To further verify the effectiveness of HRPEO, this experiment compares it with several advanced algorithms. The selected comparison algorithm and parameter settings are as follows:

DHNDE[38], 2022, $F = 0.9$, $Cr = 0.3$, $m = 5$, $dmin = 0.1$, $fmin = 1.0E - 5$, $Np = 200$, $maxio = 200$.

DQNMMA[15], 2024, $Np = 90$, $\propto = 0.001$, $\gamma = 0.9$, $\varepsilon = 0.6$, $pL = 0.2$.

EDA_ACO_LS[39], 2024, $N = 400$, $Ps = 0.1$, $\lambda^{eda} = 0.8$, $\xi^{aco} = 0.4$, $\omega = 0.4$,

EMOEA[40], $N = 60$, $CR = 0.8$, $MR = 0.5$, $r = 0.2$.
LMEO[41], 2023, $N1 = 30$, $N2 = 15$, $N3 = 15$.
PSOSA[6], 2023, $T1 = 100$, $R = 50$, $\omega = 1/(2 + log2)$, $c1 = 0.5 + log2$, $c2 = 0.5 + log2$;

we used the Friedman analysis of variance method to analyze the data of each group at the 0.05 level, and the analysis results. The experimental results are shown in Table III and Table IV. In addition, we visualized the experimental data, as shown in the box plot in Fig 13. In Fig 13, the box of HRPEO is the flattest, the data distribution is denser, and the mean of its distribution is smaller than that of some other comparison algorithms. Combining the results in Table III and Table IV, HRPEO performs better than other algorithms on 38 instances, so we conclude that HRPEO is easier to find a better solution when solving the FJSPMA problem than other comparison algorithms. Compared with HRPEO, EDA_ACO_LS and EMOEA lack exploration capabilities and local implementation strategies for AGV transportation tasks. Therefore, they are better than the other four comparison algorithms that lack search capabilities, but the experimental results are not as good as HRPEO. DQNMMA and LMEO lack local search for machine selection, and the results of the algorithms are unevenly distributed. There are data gaps below the mean, and the algorithms are greatly affected by the instance scale. Through analysis, we found that HRPEO is more able to balance the exploitation and exploration capabilities of the algorithm, and performs local search based on the characteristics of the problem, making it more suitable for solving FJSPMA.

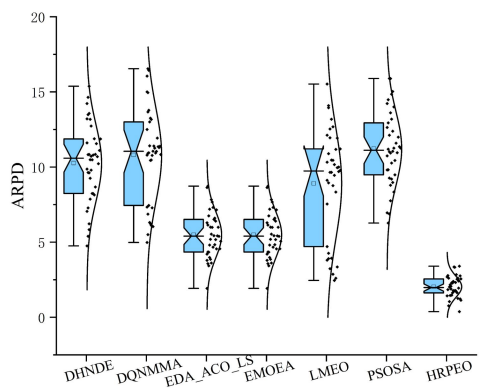

**Fig. 13.** The box plot of comparison algorithm.

## VI. CONCLUSIONS

It is a promising direction to use the strategy of solving multimodal problems to design algorithms for solving shop scheduling problems. This paper proposes a novel HRPEO algorithm to solve the flexible job shop scheduling problem with limited multi-load AGVs. This study analyzes the FJSPMA, decomposes the problem into four sub-problems, establishes the coupling relationship between the sub-problems, and designs encoding and decoding rules. A regional division strategy is designed in the HRPEO, which can not only guide the exploitation of HRPEO based on historical solutions, but also continuously explore unknown regions to prevent the algorithm from falling into local optimality. In addition, a local search is designed to improve the search efficiency of the HRPEO based on the transportation characteristics of multi-load AGVs. Finally, compared with multiple advanced algorithms, and a large number of experiments are carried out to test the performance of the HRPEO. The results show the superiority of HRPEO in solving FJSPMA.

In the future, we will continue to study the job shop scheduling problem of coupled processing and transportation. We can introduce more realistic constraints, such as path conflicts, AGV transportation speed and energy consumption. At the same time, applying multimodal problem-solving strategies to improve algorithms for solving shop scheduling problems still has significant room for improvement, such as in distinguishing individual characteristics, constructing subpopulations, and optimizing clustering methods.